\documentclass[sigconf,screen,nonacm]{acmart}

\setcopyright{none}
\renewcommand\footnotetextcopyrightpermission[1]{}
\acmConference[arXiv]{arXiv preprint}{October 2026}{}
\acmYear{2026}
\acmDOI{}
\acmISBN{}

\usepackage{amsmath}
\usepackage{algorithm}
\usepackage{algpseudocode}
\usepackage{booktabs}
\usepackage{graphicx}
\usepackage{microtype}
\usepackage{xurl}

\hypersetup{
  hidelinks,
  pdftitle={Evidence-Bound Gateway-Path Provenance for Third-Party LLM Inference},
  pdfauthor={Fei Wang, Zebai Tian},
  pdfkeywords={LLM gateway, cloud service mediation, trusted execution environment, remote attestation}
}

\newcommand{\amr}{AMR}
\newcommand{\hash}{\mathsf{H}}
\newcommand{\micros}{\ensuremath{\mu\mathrm{s}}}

\newtheorem{theorem}{Theorem}
\newtheorem{proposition}[theorem]{Proposition}
\newtheorem{lemma}[theorem]{Lemma}
\newtheorem{corollary}[theorem]{Corollary}

\title{Evidence-Bound Gateway-Path Provenance for Third-Party LLM Inference}

\author{Fei Wang}
\affiliation{%
  \institution{}
  \country{}}
\email{research@wangfei.dev}

\author{Zebai Tian}
\authornote{Corresponding author.}
\affiliation{%
  \institution{}
  \country{}}
\email{seven0709@foxmail.com}

\begin{document}

\begin{abstract}
Third-party LLM gateways select providers, models, endpoints, and fallback paths while also authoring the records used to explain and bill those decisions. We present evidence-bound gateway-path provenance, which makes these decisions subject to client-verifiable request contracts. An attested mediation runtime intersects a signed service policy with the client's authorization, constructs provider-facing calls, and protects streamed responses. The contract distinguishes initial routes from fallback destinations and admits each transition only after an authorized failure. A signed close-out binds the execution trace, provider-reported usage, authorized pricing, and complete response. We formalize the protocol state machine and establish authorization preservation, execution and stream binding, charge integrity, and channel confidentiality relative to the public receipt metadata. The analysis also characterizes replay, runtime replacement, and the information disclosed by public payload commitments. Controlled software-path measurements show 0.442~ms paired p95 latency overhead at concurrency one, peak throughput of 10.65 thousand requests/s, and stream close-out below 0.4~ms p95. An AWS Nitro Enclaves prototype completed 30 live single-route streaming requests with attestation-bound session and signing keys. The design separates gateway administration from payload processing and supplies a verifiable account of each confirmed execution.
\end{abstract}


\maketitle

\section{Introduction}

LLM applications commonly use a gateway instead of integrating separately with each LLM provider API. Cloud services such as OpenRouter, LiteLLM, Cloudflare AI Gateway, and Portkey normalize heterogeneous APIs and centralize provider selection, fallback, streaming, usage tracking, and reliability controls~\cite{openrouter_routing,openrouter_fallback,litellm_gateway,cloudflare_ai_gateway,portkey_gateway}. This consolidation makes the gateway a multi-provider service broker: for each request, it chooses not only where bytes travel, but which provider/model service is invoked, which failure permits a fallback, which endpoint is contacted, and which usage record determines the charge.

The resulting trust boundary is problematic. Application-facing TLS usually terminates at the gateway, exposing prompts, tool calls, credentials, and responses to the operator. The same operator also controls the mediation decisions and their logs. A malicious or misconfigured gateway can therefore substitute a cheaper provider or model, bypass an initial route, hide a fallback, rewrite an endpoint, alter a response, or fabricate usage while returning a normal API response. These are practical risks rather than hypothetical API irregularities. A measurement of 428 commodity LLM routers found malicious code injection into returned tool calls, access to canary credentials, and one transfer from a researcher-controlled wallet~\cite{liu2026agentmine}. Audits of third-party LLM APIs also report deceptive model claims and behavior inconsistent with advertised services~\cite{zhang2026realmoney,pasquini2025llmmap}.

Trusted execution environments (TEEs) can isolate gateway code and support remote attestation~\cite{sabt2015tee,sok2022tee,rfc9334}. However, a useful cloud gateway cannot freeze every routing choice into one measured image. Available providers, client permissions, fallback conditions, endpoints, and prices change more frequently than deployed code. The central systems problem is therefore to combine a stable, attestable enforcement mechanism with authenticated request-specific semantics. Measuring a relay or gateway binary establishes which program is running; it does not by itself identify the late-bound contract that this request authorized or the route trace that the program produced.

We address this problem with a \emph{contract-aware attested semantic data plane}. The gateway business plane retains authentication, quotas, instance selection, opaque forwarding, and record storage. An attested mediation runtime (\amr{}) inside a TEE terminates a client-to-runtime protected channel, intersects an operator service policy with a client contract, admits provider/model routes and failure-guarded transitions, validates provider-facing endpoints, normalizes provider-reported usage under a client-authorized pricing policy, and protects streamed responses. A signed close-out record lets the client confirm which permitted execution occurred and binds that execution to the complete response.

This paper makes three contributions:
\begin{itemize}
  \item It formulates multi-provider LLM gateway routing as \emph{request-scoped cloud service mediation}. The contract separates initial routes from fallback destinations and binds fallback guards, endpoint constraints, an attempt bound, and a pricing-policy commitment.
  \item It specifies and implements a split-plane gateway, and develops a compositional security analysis connecting authorized execution, authenticated streaming, charge evidence, and runtime lifecycle. The confidentiality result explicitly accounts for metadata exported in public receipts.
  \item It characterizes the latency, throughput, and response-shape dependence of the complete protected path, and demonstrates live streaming through an AWS Nitro Enclaves prototype. A scenario-based validation appendix connects the protocol checks to retained execution records.
\end{itemize}

\section{Multi-Provider Cloud Mediation}
\label{sec:model}

\subsection{System and Adversary}

An invocation involves a client, a gateway business plane, an \amr{}, and an LLM provider. The client states the service conditions under which it will accept the result. The business plane exposes the public gateway endpoint and performs administrative functions, but forwards protected request and response frames without terminating their encryption. The \amr{} alone decrypts the request, constructs the provider-facing LLM API call, and authenticates the mediation outcome. The provider remains outside the gateway trust boundary.

We give the business plane full control over gateway-visible communication and stored records. It may delay, drop, replay, reorder, or substitute messages; direct a new session to any available runtime instance; and suppress its close-out record. The authorized \amr{} and client verifier form the software trusted computing base. We assume standard signature, hash, and authenticated-encryption security, sound remote attestation, and the selected TEE threat model. Availability, traffic-shape confidentiality, a compromised client or provider, and physical or microarchitectural attacks outside that model are out of scope.

The client pins a registry verification key and maintains a client-side set of acceptable runtime measurements. A signed registry advertises available measurements and the service-policy verification key, but cannot extend the client's measurement policy. Thus the operator can publish or withdraw an implementation, while the client decides which measured contract interpreter it trusts.

\subsection{Request-Scoped Contract}

A route is a concrete provider-facing service path
\begin{equation}
  r=(p,m,\alpha,e),
\end{equation}
where $p$ is an LLM provider, $m$ is its model identifier, $\alpha$ identifies the provider account, and $e$ constrains the TLS-authenticated endpoint by scheme, host, port, and path prefix. The account identifier names a credential binding without carrying its secret. A request contract is
\begin{equation}
  \mathcal{C}=(R_0,R,F,k,h_\Pi).
  \label{eq:contract}
\end{equation}
$R$ contains admitted routes, nonempty $R_0\subseteq R$ contains routes from which execution may begin, and the positive integer $k$ bounds the number of attempts. Let $\mathcal{W}_f$ be the failure outcomes classified by the runtime's provider adapter. Each edge $(u,\Omega,v)\in F$, with $u,v\in R$ and $\varnothing\ne\Omega\subseteq\mathcal{W}_f$, permits fallback from $u$ to $v$ only when the preceding outcome belongs to $\Omega$. Success is not a failure outcome. The commitment $h_\Pi=\hash(\Pi)$ identifies a client-authorized pricing policy $\Pi$ that maps provider/model and token classes to exact decimal rates.

The operator signs a service policy $\mathcal{P}$ in the same language. The \amr{} computes the effective contract $\mathcal{C}_{\mathrm{eff}}=\mathcal{P}\sqcap\mathcal{C}$: route and initial-route sets intersect, fallback edges remain only when both policies admit the transition and failure guard, the attempt bound takes the smaller value, and pricing-policy commitments must agree. The meet fails if any mandatory component is inconsistent. Route identity includes its provider, model, account selection identifier, and endpoint constraint; the implementation retains matching route definitions rather than merging conflicting endpoint strings.

For an explicit interpretation, let $T_X$ expand the guarded edges of contract $X$ into triples $(u,\omega,v)$, one per admitted failure outcome. Then
\begin{equation}
\begin{aligned}
 R_{\mathrm{eff}} &= R_{\mathcal P}\cap R_{\mathcal C},\\
 R_{0,\mathrm{eff}} &= R_{0,\mathcal P}\cap R_{0,\mathcal C}\cap R_{\mathrm{eff}},\\
 T_{\mathrm{eff}} &= \{(u,\omega,v)\in T_{\mathcal P}\cap T_{\mathcal C}
                    \mid u,v\in R_{\mathrm{eff}}\},\\
 k_{\mathrm{eff}} &= \min(k_{\mathcal P},k_{\mathcal C}).
\end{aligned}
\label{eq:narrowing}
\end{equation}
The effective edge set $F_{\mathrm{eff}}$ groups the triples in $T_{\mathrm{eff}}$ by source and destination, collecting their outcomes into guards. The operation rejects unequal pricing commitments or empty route or initial-route sets. An empty effective edge set disables fallback while retaining authorized initial routes.

An execution yields a nonempty route trace $\rho=\langle a_0,\ldots,a_j\rangle$. Attempt $a_i=(i,r_i,\hat e_i,\omega_i)$ records its sequence, route, endpoint observation, and terminal outcome. A failure before TLS establishment has $\hat e_i=\bot$; a successful attempt requires an authenticated endpoint observation. For a contract $X$, write $\rho\models X$ when (i) $r_0\in R_{0,X}$; (ii) each $r_i\in R_X$ and each authenticated endpoint observation satisfies its route's constraint; (iii) $j+1\leq k_X$; (iv) for every $i>0$, some $(r_{i-1},\Omega,r_i)\in F_X$ contains $\omega_{i-1}$ in its guard; and (v) $\omega_j=\textsf{success}$ and every earlier outcome belongs to $\mathcal{W}_f$. Applying this predicate to $X=\mathcal{C}_{\mathrm{eff}}$ defines confirmed routing compliance. If $A$ is the required primary and $B$ is enabled only after a retriable failure at $A$, such as a timeout, selecting $B$ directly violates the initial-route constraint.

Table~\ref{tab:example} illustrates the distinction. Its first four traces use only the admitted routes $A$ and $B$, yet differ in initial-route and transition compliance. The final trace uses $B'$, which retains $B$'s provider and model but changes its endpoint, illustrating the separate endpoint constraint. Destination membership determines which routes are available; guarded transitions determine when they may be used.

\begin{table}[t]
\caption{Example contract: start at $A$; use $B$ only after a retriable failure at $A$.}
\label{tab:example}
\centering
\footnotesize
\setlength{\tabcolsep}{3.1pt}
\begin{tabular}{lll}
\toprule
Observed trace & Decision & Reason \\
\midrule
$A{:}\textsf{success}$ & Accept & Initial route succeeds \\
$A{:}\textsf{timeout},B{:}\textsf{success}$ & Accept & Guarded edge enabled \\
$B{:}\textsf{success}$ & Reject & $B\notin R_0$ \\
$A{:}\textsf{terminal},B{:}\textsf{success}$ & Reject & Failure guard not met \\
$A{:}\textsf{timeout},B'{:}\textsf{success}$ & Reject & Endpoint mismatch \\
\bottomrule
\end{tabular}
\end{table}

\subsection{Required Properties}

The design targets four properties. \emph{Authorized execution} requires every provider-facing attempt to extend an authorized prefix. \emph{Execution and stream binding} requires a confirmed result to belong to the retained request and match the complete client-observed response. \emph{Charge integrity} binds the successful route and runtime-observed provider usage to the authorized pricing policy. \emph{Channel confidentiality} protects payloads beyond the metadata intentionally exported by the protocol. Section~\ref{sec:security} defines these properties and the leakage of the public-receipt interface.

Attestation identifies the interpreter and its keys. The authenticated contract supplies request-specific authorization, and the trace records its execution. This separation lets one measured program enforce different client policies while preserving a verifiable account of each request.

\begin{figure*}[!t]
  \centering
  \includegraphics[width=0.84\textwidth]{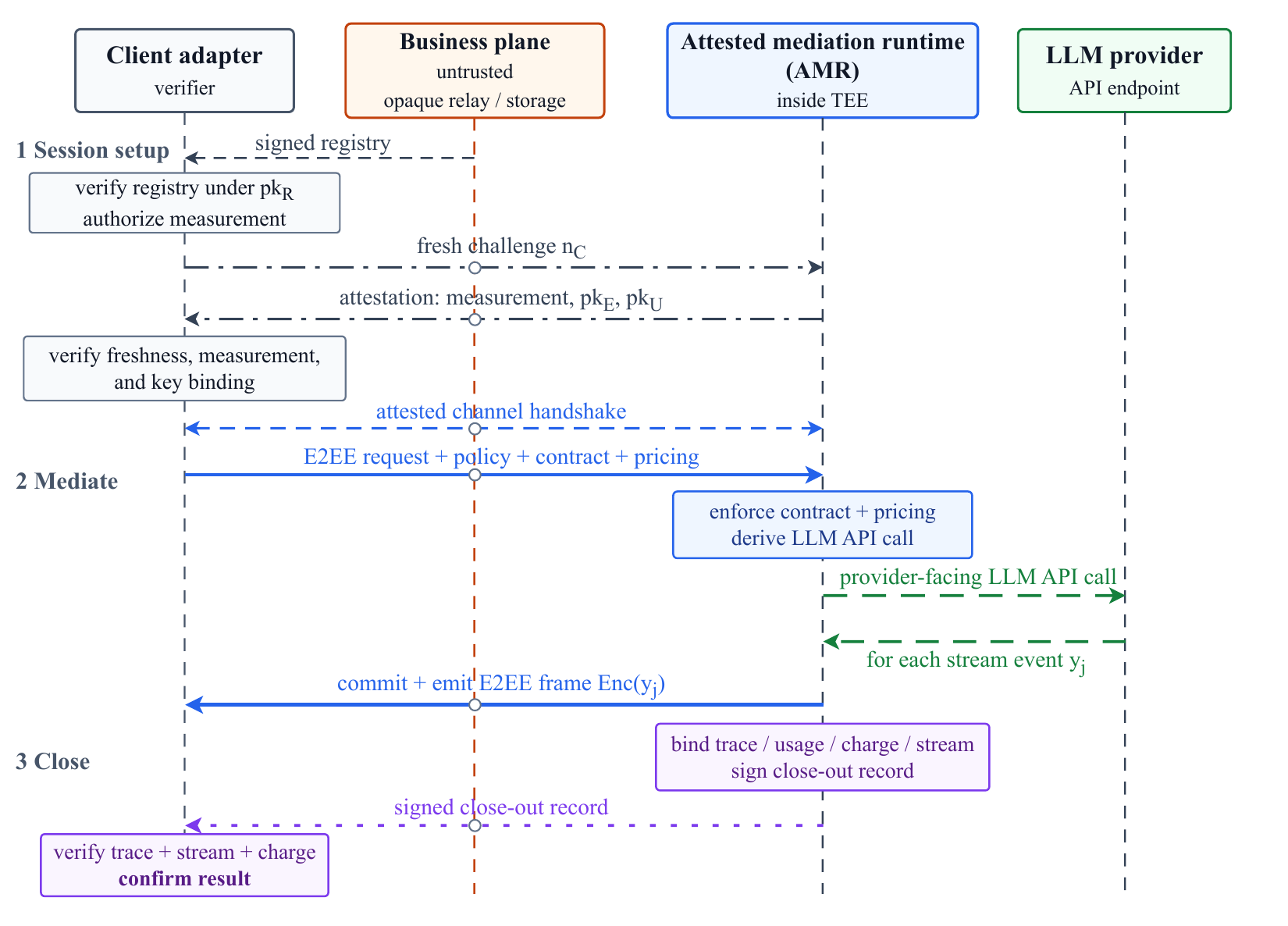}
  \caption{Attested mediation lifecycle. Client--AMR payloads cross the business plane as opaque ciphertext. Each provider event is committed and forwarded immediately as an encrypted frame; stream close produces the signed close-out record.}
  \Description{Sequence diagram showing registry and attestation verification, an encrypted client--runtime channel, runtime--provider TLS, incremental protected responses, and signed close-out verification.}
  \label{fig:architecture}
\end{figure*}

\section{Contract-Aware Attested Architecture}
\label{sec:design}

Figure~\ref{fig:architecture} shows the three protocol phases and four trust domains. The placement rule is simple: the component that sees mediated plaintext must also interpret the effective contract and construct the provider-facing request. Otherwise the business plane could change semantics before or after protected processing.

\subsection{Attested Session Establishment}

The client obtains the signed registry through the business plane or another distribution path. It verifies the registry under its pinned key and filters advertised measurements through its client-side measurement policy. For a selected runtime, the client sends a fresh challenge $n_C$. Attestation evidence binds $n_C$, the runtime measurement $\delta$, a channel key $pk_E$, and a distinct close-out signing key $pk_U$ to one runtime context~\cite{coker2011principles,rfc9334}. The client validates the evidence chain, freshness, platform state, and $\delta$ membership before completing an authenticated key exchange with $pk_E$. Request material is released only through the resulting protected channel.

Attestation intentionally binds keys to the measured \emph{interpreter}, not to one static route list. The signed operator policy, client contract, and pricing policy arrive after session establishment as authenticated channel inputs. This separation permits routing and prices to change without rebuilding the TEE image, while measurement changes still require explicit client authorization.

\subsection{In-Runtime Mediation}

After decryption, the \amr{} verifies the service-policy signature, validates the client contract, computes $\mathcal{C}_{\mathrm{eff}}$, and checks $\hash(\Pi)=h_\Pi$. It rejects before network access if the meet is undefined or a required rate is absent. For each attempt, the runtime chooses an admitted initial route or a fallback enabled by the preceding outcome, derives the provider-specific API body and headers, constructs the endpoint from the selected route, and validates TLS service identity~\cite{rfc9525}. The attempt result extends $\rho$; a success closes routing, whereas a permitted failure may activate one outgoing edge.

\begin{figure*}[t]
  \centering
  \includegraphics[width=0.97\textwidth]{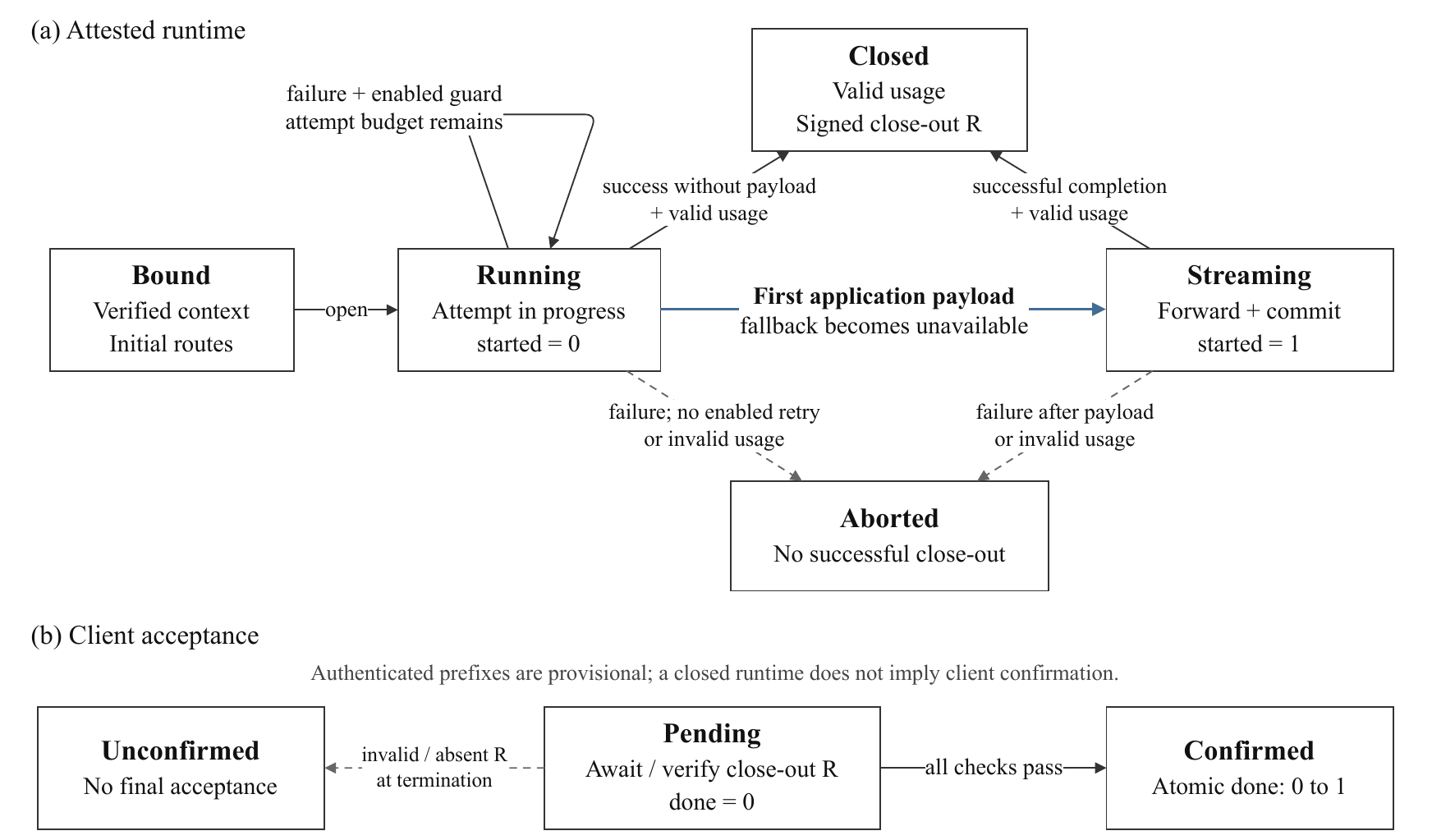}
  \caption{Runtime execution and client acceptance. Before the first application payload, a classified failure may enable another authorized attempt within the attempt bound. Emitting a payload disables fallback. Successful closure produces a signed record $R$; final client acceptance requires verification and an atomic one-time confirmation. Dashed arrows denote failure paths.}
  \Description{Two-panel state machine. The runtime moves from Bound to Running, may retry after an authorized failure before any payload, and enters Streaming on its first application payload. Valid successful completion leads to Closed; an unrecoverable failure or invalid usage leads to Aborted. Separately, the client moves from Pending to Confirmed only after all close-out checks pass, or to Unconfirmed if the record is invalid or absent at termination.}
  \label{fig:mediation-state-machine}
\end{figure*}

Algorithm~\ref{alg:mediate} gives the semantic state machine using the components of $\mathcal{C}_{\mathrm{eff}}$. Figure~\ref{fig:mediation-state-machine} summarizes its runtime transitions and the separate client acceptance step. The candidate set $A$ begins with its initial routes and is updated after a failure using the enabled edges. Endpoint admission precedes network access. The opening operation validates TLS service identity before transmitting $D(Q,r)$ and returns either a stream with status \textsf{open} or a classified failure. Its authenticated endpoint observation is retained as evidence. Stream closure supplies the terminal outcome and provider-reported usage $\mu$.

\begin{algorithm}[t]
\caption{Contract-constrained mediation}
\label{alg:mediate}
\footnotesize
\begin{algorithmic}[1]
\Require request $Q$, signed policy $\mathcal{P}$, contract $\mathcal{C}$, pricing $\Pi$
\State verify $\mathcal{P}$; $\mathcal{C}_{\mathrm{eff}}\gets\mathcal{P}\sqcap\mathcal{C}$
\State reject if $\mathcal{C}_{\mathrm{eff}}=\bot$
\State $(R_0,R,F,k,h_\Pi)\gets\mathcal{C}_{\mathrm{eff}}$
\State reject if $\hash(\Pi)\neq h_\Pi$; validate rates for $R$
\State $A\gets R_0$; $\rho\gets\langle\rangle$
\For{$i=0,\ldots,k-1$}
  \State $r\gets\mathsf{Select}(A)$; verify $r$ and its endpoint constraint
  \State $started\gets\mathsf{false}$; $s\gets\mathsf{InitCommit}()$
  \State $(S,\hat e,\omega)\gets\mathsf{Open}(r,\mathsf{Derive}(Q,r))$
\If{$\omega=\textsf{open}$}
    \State verify authenticated endpoint $\hat e$
    \ForAll{content event $y_j$ from $S$}
      \State $s\gets\mathsf{Update}(s,y_j)$; emit authenticated frame $y_j$
      \State $started\gets\mathsf{true}$
    \EndFor
    \State $(\omega,\mu)\gets\mathsf{Close}(S)$
  \EndIf
  \State $\rho\gets\rho\mathbin{\|}(i,r,\hat e,\omega)$
  \If{$\omega=\textsf{success}$}
    \State validate $\mu$; $\beta\gets\mathsf{Bill}(\Pi,r,\mu)$
    \State \Return signed close-out $(\rho,\mu,\beta,s)$
  \EndIf
  \State reject if $started$
  \State $A\gets\{r'\mid(r,\Omega,r')\in F\land\omega\in\Omega\}$
  \State reject if $A=\varnothing$
\EndFor
\State \Return reject
\end{algorithmic}
\end{algorithm}

Provider response events are transformed into authenticated client--\amr{} frames as they arrive. A content event in Algorithm~\ref{alg:mediate} is any application-visible response payload, including tool and application metadata events; transport handshakes and control records are separate. The runtime never concatenates application payloads across routes: fallback is attempted only before such a payload is emitted. A later stream failure terminates without a close-out. The runtime updates a domain-separated SHA-256 commitment over the pre-encryption application bytes; the client updates the same commitment after decryption. Prefixes remain incrementally available, while changes to the delivered byte sequence affect final confirmation.

\subsection{Usage Normalization and Close-Out}

Let the successful route be $(p^*,m^*,\alpha^*,e^*)$, and let $\mu$ denote its provider-reported usage, normalized to input, cached-input, and output counts $(n_i,n_c,n_o)$. With client-authorized per-million rates $(\lambda_i,\lambda_c,\lambda_o)$ from $\Pi$, the runtime computes the expected charge
\begin{equation}
  \beta=\frac{\lambda_i(n_i-n_c)+\lambda_cn_c+\lambda_on_o}{10^6}.
  \label{eq:bill}
\end{equation}
The runtime requires $0\leq n_c\leq n_i$, nonnegative rates and counts, a matching provider/model rate, and representable exact-decimal arithmetic. Missing usage or an invalid input prevents a charge-bearing close-out. The charge is bound to the provider's reported usage and the authorized policy $\Pi$.

At stream close, the \amr{} signs a compact record containing commitments to the registry, policies, effective contract, pricing policy, and request; the complete trace $\rho$; provider-reported usage; expected charge; response byte count, completion marker, and stream commitment; the challenge; and $pk_U$. A client confirms the record by verifying $pk_U$ against the attested session, recomputing the policy meet and request/stream commitments, checking $\rho\models\mathcal{C}_{\mathrm{eff}}$, and recomputing Eq.~\eqref{eq:bill}. Response prefixes may be consumed provisionally, but a missing or invalid close-out cannot confirm the complete mediation result.

\section{Security Model and Analysis}
\label{sec:security}

\subsection{Adversary, Interfaces, and Assumptions}
\label{sec:security-assumptions}

Let $\kappa$ be the security parameter. The adversary is probabilistic polynomial-time and controls the business plane, parent host, message scheduling, and retained records. It can interleave sessions, submit its own requests, select operator policies, induce network failures, redirect traffic, and start replicas of approved images. It cannot read an honest enclave's memory, change the client's retained authorization state, or compromise the credential owner or provider in the experiment. An operator policy may be restrictive or adversarial: its signature establishes origin, while the client contract limits its authority.

The analysis composes five interfaces. Attestation binds a fresh challenge, approved measurement, and runtime keys. An authenticated key exchange binds the client to that runtime. Ordered authenticated channels carry inference and credential payloads; provider TLS authenticates the admitted service. A typed serializer and domain-separated commitments bind structured values and response bytes. Finally, the runtime signs close-out records under its attested signing key. The Noise and TLS instantiations supply the channel interfaces~\cite{noiseprotocol,rfc9525}; the attestation interface follows the challenge and key-binding model of Section~\ref{sec:design}~\cite{coker2011principles,rfc9334}.

The serializer gives each accepted object one interpretation at both endpoints. Distinct types and protocol roles use distinct commitment and signature domains. We write $H_d(x)$ for a commitment under domain $d$; the earlier notation $\hash(\Pi)$ abbreviates the pricing-policy domain. Collision resistance applies to the encoded values. It supplies binding, not secrecy of a publicly exported digest.

The channel abstraction authenticates the session, request, direction, and record order before delivering application bytes. These labels can be explicit authenticated fields or implicit in separate per-request channel state. We use multi-session channel-realization guarantees: original application messages are delivered privately while observable handshake, length, and scheduling events admit an online simulation. Each honest client uses a fresh challenge for a new attested session, unique request identifiers within that session, and a terminal acceptance flag for each request. Client trust state and release-signing keys are outside the corruption interface. Valid operator policies remain adversarially selectable.

For all polynomially bounded sessions and messages, let $\epsilon_{\rm att}$, $\epsilon_{\rm reg}$, $\epsilon_{\rm ake}$, $\epsilon_{\rm rec}$, $\epsilon_{\rm tls}$, $\epsilon_{\rm sig}$, and $\epsilon_H$ bound failures of attestation binding, registry authentication, key exchange, record authentication, provider TLS authentication, close-out signature unforgeability, and commitment binding. They bound the corresponding bad events over all sessions in each relevant experiment, rather than a single operation. Let $\epsilon_{\rm uniq}$ bound accidental context reuse or key collisions between fresh incarnations. For $\nu$ independent $b$-bit identifiers in one namespace, its collision contribution is at most $\nu(\nu-1)/2^{b+1}$. Define
\begin{equation}
\begin{aligned}
 \epsilon_{\rm auth}={}&\epsilon_{\rm att}+\epsilon_{\rm reg}+\epsilon_{\rm ake}
                         +\epsilon_{\rm rec}\\
                       &+\epsilon_{\rm tls}+\epsilon_{\rm sig}
                         +\epsilon_H+\epsilon_{\rm uniq}.
\end{aligned}
\label{eq:auth-budget}
\end{equation}
The proofs first condition on none of these events occurring, denoted $\mathsf{Good}$, and then apply the union bound. With secure primitives and fresh contexts, $\epsilon_{\rm auth}(\kappa)$ is negligible.

\subsection{Request Context and Protocol Machine}
\label{sec:protocol-machine}

Let $\mathsf{sid}$ identify one authenticated channel instance, binding the accepted challenge, attested runtime keys, and authenticated handshake context. A peer key and a caller-supplied nonce alone do not identify a distinct channel instance. Before releasing a request, the client retains
\begin{equation}
 \mathcal{S}_q=(\mathsf{sid},\mathsf{rid},Q,G,\mathcal{P},\mathcal{C},\Pi,\mathsf{done}),
\label{eq:retained-state}
\end{equation}
where $\mathsf{done}=0$ initially and $G$ is the verified registry snapshot. The request identifier and channel identify this state throughout execution. The client derives the expected commitments from these retained objects, not from replacement objects supplied alongside a receipt. Loss of the live acceptance state requires a fresh context; a stored receipt alone cannot restore a pending transaction.

Let $D(Q,r)$ be the committed semantic request derivation. For the initial route $r_0$, define $d_Q=H_Q(D(Q,r_0))$. Verification recomputes this derivation from the retained request and the initial route in the validated trace. Any request-specific parameters needed by $D$ belong to the authenticated context. The close-out binds
\begin{equation}
 \gamma_q=(\mathsf{sid},\mathsf{rid},h_G,h_{\mathcal P},h_{\mathcal C},
             h_{\rm eff},h_\Pi,d_Q).
\label{eq:receipt-context}
\end{equation}
This is a logical context: every execution-distinguishing component must be covered by signed fields or a signed context commitment and checked against the retained state. Merely storing a value locally does not bind it to a receipt. During binding $d_Q$ is unset; selecting the initial route fixes it once for the request. The derivation $D$ need not be injective: the original $Q$ is authenticated by its unique request/channel context, while $d_Q$ binds the derived call.

For reasoning about Algorithm~\ref{alg:mediate}, write a runtime state as
\begin{equation}
 \Sigma_q=(\gamma_q,\mathsf{phase},A,\rho,r_{\rm cur},Y,\mathsf{started}).
\label{eq:runtime-state}
\end{equation}
$Y$ is the history of emitted application bytes, used as proof notation; the implementation maintains an incremental digest rather than buffering $Y$. The states are \textsf{bound}, \textsf{running}, \textsf{streaming}, \textsf{closed}, and \textsf{aborted}. Binding authenticates the inputs and computes the effective contract. Opening a route records a pending attempt $r_{\rm cur}$. Stream events extend $Y$. A classified failure completes an attempt and may enable the next route only while $\mathsf{started}=0$. Successful closure validates usage, computes the charge, and signs exactly one terminal record. An abort emits no successful close-out. Closed and aborted states have no further provider-call transitions for that request.

Define $\mathsf{Pref}_X$ as the attempted histories satisfying the initial-route, endpoint, transition, and attempt-bound constraints of contract $X$, without requiring final success. An active attempt may have a pending outcome; every transition to a subsequent attempt requires a completed failure at its predecessor. A successful closed history belongs to $\mathsf{Pref}_X$ and additionally satisfies $\rho\models X$.

\subsection{Authorization Preservation}

\begin{proposition}[Policy intersection narrows authorization]
\label{prop:narrowing}
For $\mathcal{C}_{\rm eff}=\mathcal{P}\sqcap\mathcal{C}\ne\bot$,
\begin{equation}
 \mathsf{Pref}_{\mathcal{C}_{\rm eff}}
 \subseteq\mathsf{Pref}_{\mathcal P}\cap\mathsf{Pref}_{\mathcal C}.
\label{eq:prefix-inclusion}
\end{equation}
Every closed trace satisfying $\mathcal{C}_{\rm eff}$ therefore satisfies both input contracts. The effective pricing commitment equals both input commitments.
\end{proposition}
\begin{proof}
Every effective initial route and route belongs to both inputs by Eq.~\eqref{eq:narrowing}. Every effective transition triple belongs to both transition relations, so it preserves its predecessor, destination, and enabling outcome. Matching route definitions preserve endpoint constraints. The effective attempt bound is no larger than either input bound. These componentwise inclusions establish Eq.~\eqref{eq:prefix-inclusion}, including histories with a pending final attempt. Final success is the same closure condition in each contract. Pricing commitments must agree for the intersection to be defined.
\end{proof}

\begin{lemma}[Authorized-prefix invariant]
\label{lem:prefix}
Every reachable state of Algorithm~\ref{alg:mediate} has an issued-call history in $\mathsf{Pref}_{\mathcal{C}_{\rm eff}}$. All emitted content belongs to at most one route attempt. A successful close-out contains exactly the completed issued-call trace and its successful final attempt.
\end{lemma}
\begin{proof}
The invariant holds for the empty history after binding. For the first opening, $A=R_{0,\rm eff}$, and endpoint admission precedes network access. A pending attempt therefore extends an authorized history. Completing an attempt records its actual endpoint evidence and classified outcome without changing the route already issued. If it fails before content emission, the next candidate set contains exactly the destinations enabled by that failure. If it fails after content emission, the request aborts. Thus no second attempt can contribute content after $\mathsf{started}$ becomes true. Each opening consumes one loop iteration, giving the attempt bound. A success returns immediately after usage validation and signing; it cannot be followed by another call. By induction over these transitions, the invariant and trace correspondence hold at every reachable state.
\end{proof}

\begin{theorem}[Authorized provider invocation]
\label{thm:execution}
Under $\mathsf{Good}$, an honest runtime processing an honest client's request issues only calls authorized by both $\mathcal P$ and $\mathcal C$. Each established provider connection satisfies its admitted endpoint constraint, and the provider-facing semantic request is $D(Q,r)$ for the issued route $r$.
\end{theorem}
\begin{proof}
Channel authentication binds $Q$ and the contract inputs to the retained request context. The signed policy is authenticated, and intersection cannot enlarge client authorization by Proposition~\ref{prop:narrowing}. Lemma~\ref{lem:prefix} then applies to every issued call. The only call transition constructs $D(Q,r)$ and checks the destination before releasing request material; provider TLS authenticates the resulting service. A retry or redirect that retransmits the inference request must be represented as another admitted attempt, subject to the same guard, endpoint, trace, and attempt-bound checks. Hence a host-induced address change either satisfies the constraint or causes failure. Network delay can enable an authorized failure guard, but it cannot create a new edge or justify bypassing the initial route.
\end{proof}

The theorem concerns runtime-issued calls at authenticated service boundaries. It does not infer the provider's internal model implementation from a model name in the API request.

\subsection{Close-Out Authenticity and Context Binding}
\label{sec:record-binding}

Write a close-out payload as
\begin{equation}
 B_q=(\gamma_q,\rho,\mu,\beta,d_Y,\ell,\mathsf{complete},pk_U),
\label{eq:close-out}
\end{equation}
where $d_Y=H_{\rm stream}(Y)$ and $\ell=|Y|$. Its signature covers a domain-tagged encoding of the complete payload. Verification and a new live confirmation are distinct operations:
\begin{equation}
\begin{aligned}
 \mathsf{VerifyResult}={}&\mathsf{KeyOK}\land\mathsf{ContextOK}\land\mathsf{TraceOK}\\
                       &\land\mathsf{BillOK}\land\mathsf{StreamOK},\\
 \mathsf{Confirm}={}&\neg\mathsf{done}\land\mathsf{VerifyResult}.
\end{aligned}
\label{eq:confirm}
\end{equation}
$\mathsf{KeyOK}$ verifies the signature under the key from the accepted attestation. $\mathsf{ContextOK}$ compares Eq.~\eqref{eq:receipt-context} with commitments recomputed from $\mathcal S_q$. $\mathsf{TraceOK}$ validates the complete trace under the recomputed effective contract. $\mathsf{BillOK}$ recomputes the charge and checks its currency label. $\mathsf{StreamOK}$ checks completion, byte count, and the digest of the client's own decrypted bytes. The latter state is computed from received content, not copied from the close-out. A new confirmation atomically checks $\mathsf{done}=0$ and sets it to one after successful verification.

\begin{lemma}[Issued-record and request correspondence]
\label{lem:record}
Under $\mathsf{Good}$, a confirmed payload was signed by the runtime identified in the accepted session for the same retained request and authorization context. One request context cannot be confirmed twice.
\end{lemma}
\begin{proof}
If the payload had not been issued under the accepted signing key, its valid signature would constitute a signature forgery. Attestation and authenticated key exchange associate that key with the selected runtime. A payload from another channel instance or request fails the signed context comparison, except upon context reuse or commitment failure. Within that context, record authentication fixes the original request supplied to the runtime. Different retained policy objects fail their commitment comparisons except upon a hash collision; the derived-call commitment is checked using that authenticated request and the initial route. These events are excluded by $\mathsf{Good}$. The acceptance flag prevents a second confirmation even when the same valid payload is replayed. This proves correspondence without requiring randomized signatures to be unique.
\end{proof}

Stored-record rechecking evaluates $\mathsf{VerifyResult}$ with the original authorization and locally recorded stream state; it does not reset $\mathsf{done}$ or create another live confirmation. A later registry or price update does not reinterpret an earlier signature or recompute its historical charge under new terms.

\subsection{Streaming Authenticity and Completion}

Let $Y_A$ be the application-byte sequence emitted by the runtime for a request and $Y_C$ the sequence decrypted by the client before optional local API re-encoding. Equality concerns these mediated bytes, independently of packet boundaries or the provider's original wire format. An empty successful response has $Y_A=\varepsilon$.

\begin{theorem}[Authentic prefixes and complete-stream binding]
\label{thm:stream}
Under $\mathsf{Good}$, every delivered client prefix is an in-order prefix emitted for that request. If final confirmation succeeds, $Y_C=Y_A$ and the runtime closed the stream successfully. The successful final attempt is the only possible source of delivered application content.
\end{theorem}
\begin{proof}
For the prefix claim, induct on the ordered channel records accepted by the client. The first record must authenticate under the selected session and request. For each successor, direction, request binding, and record order authenticate it as the next runtime emission. An inserted, duplicated, cross-request, or reordered record cannot advance this state without breaking record authentication. Dropping a suffix leaves a valid shorter prefix.

For final confirmation, Lemma~\ref{lem:record} identifies the runtime-issued payload. Successful $\mathsf{StreamOK}$ gives $H_{\rm stream}(Y_C)=H_{\rm stream}(Y_A)$, matching byte counts and a signed completion marker. Distinct byte strings with this digest would violate commitment binding. By Lemma~\ref{lem:prefix}, content from an earlier failed attempt would force an abort; all delivered content therefore comes from the successful final attempt. This also covers the empty response. A received prefix cannot be promoted to a complete result by suppressing or replacing the close-out.
\end{proof}

An application may display an authenticated prefix immediately. It gates an irreversible action requiring a completed transaction on Eq.~\eqref{eq:confirm}. This separates incremental use from final acceptance without requiring response buffering.

\subsection{Usage and Charge Integrity}

Let $N_p(\mu)$ be the specified normalization of provider $p$'s usage object into $(n_i,n_c,n_o)$. Invalid or absent usage has no normalized value. Let $\mathsf{Bill}(\Pi,r,\mu)$ be Eq.~\eqref{eq:bill} with the rate selected by the successful route's provider and model, valid counts, and exact-decimal arithmetic in the policy's currency.

\begin{theorem}[Integrity of the confirmed charge]
\label{thm:charge}
Under $\mathsf{Good}$, a confirmed charge satisfies
\begin{equation}
 \beta=\mathsf{Bill}(\Pi,r_j,\mu),
\label{eq:confirmed-charge}
\end{equation}
where $r_j$ is the successful final route, $\Pi$ is the retained client-authorized policy, and $\mu$ is the usage observed by the runtime on that execution. An invoice accepted by exact comparison has the same currency and amount.
\end{theorem}
\begin{proof}
Lemma~\ref{lem:record} fixes the runtime-issued trace, usage, and charge. Lemma~\ref{lem:prefix} identifies its successful final route. Provider TLS and the runtime's usage parser associate $\mu$ with that execution. $\mathsf{ContextOK}$ binds the retained pricing policy and its commitment, so a different rate table cannot be substituted without a commitment collision. $\mathsf{BillOK}$ independently normalizes the bound usage and evaluates the deterministic decimal expression. It rejects absent usage, invalid counts, unavailable rates, or an amount unequal to Eq.~\eqref{eq:confirmed-charge}. Invoice comparison applies the same currency-and-amount equality at settlement.
\end{proof}

The property authenticates observed usage and its charge derivation. Provider-internal token measurement and actual transfer of funds are separate from this execution-evidence interface.

\subsection{Leakage-Aware Channel Confidentiality}
\label{sec:confidentiality}

Confidentiality must account for the public close-out, not only encrypted transport. For an execution history $\tau$, let $\mathcal L(\tau)$ export the business-plane interfaces: public registry, attestation, and handshake data; selected runtime and visible destinations; traffic lengths, directions, and times; exposed management metadata; and every public close-out field and signature. In particular, $\mathcal L$ includes route history, provider-reported usage, charge, response identifiers when exported, and request/stream commitments. Fields visible on an interface belong to the leakage whether or not they were intended to be informative.

Consider a real experiment in which the adversary observes these interfaces and the protected channel ciphertexts. In the ideal experiment, honest parties execute the same request machine, while a simulator receives $\mathcal L$ online and supplies the adversary-facing channel representation. The adversary has the same scheduling and failure controls in both experiments. Define $\mathsf{Adv}_{\mathcal L}^{\rm conf}$ as its distinguishing advantage. Let $\epsilon_{\rm chan}^{\rm conf}$ and $\epsilon_{\rm tls}^{\rm conf}$ be the multi-session confidentiality bounds for the owner/client--runtime channels and runtime--provider TLS, respectively.

\begin{theorem}[No additional channel leakage]
\label{thm:confidentiality}
For the protocol machine and leakage interface above, there is an online simulator such that
\begin{equation}
 \mathsf{Adv}_{\mathcal L}^{\rm conf}
 \leq 2\epsilon_{\rm auth}+\epsilon_{\rm chan}^{\rm conf}
                         +\epsilon_{\rm tls}^{\rm conf}.
\label{eq:confidentiality-bound}
\end{equation}
Thus protected transport exposes no efficiently distinguishable information beyond $\mathcal L$, under the stated channel assumptions.
\end{theorem}
\begin{proof}
First replace the real and ideal experiments by versions that stop upon an authentication, binding, or context-reuse bad event. Each replacement changes a distinguisher's acceptance probability by at most $\epsilon_{\rm auth}$. In the resulting games, authenticated session establishment fixes the honest channel endpoints, and an adversarial ciphertext cannot deliver a new application plaintext. Replace the honest protected records on the owner/client--runtime links, over all sessions, by the channel idealization that preserves public length and scheduling events. Its realization bound limits the change in the adversary's view by $\epsilon_{\rm chan}^{\rm conf}$. Next replace the provider TLS record representation, with change at most $\epsilon_{\rm tls}^{\rm conf}$. The honest request machine continues to receive the original application messages through these ideal channels. In particular, receipt digests, provider outcomes, and usage are computed from the original execution, not from dummy payloads.

At the final hybrid, the simulator generates the ideal channel representation using the public event trace and forwards the exact exports received from $\mathcal L$. The ideal interfaces supply authenticated public handshake information and the indicated lengths; the simulator neither guesses hidden payloads nor reconstructs their commitments. Invalid injections cannot advance application state, and the adversary's delays and drops are passed to the same scheduling interface. This yields the ideal view. The triangle inequality gives the two channel terms and the two bad-event terms in Eq.~\eqref{eq:confidentiality-bound}.
\end{proof}

\textbf{What a public commitment reveals.} Binding and hiding are different properties. Suppose the adversary knows the remaining serialization context and the request is one of two candidates $Q_0,Q_1$. From a public $d_Q$ it can compute their derived commitments and identify the matching candidate whenever those values differ. This attack uses no hash collision and does not violate transport encryption. Public stream digests similarly support candidate checking. Consequently, arbitrary-message indistinguishability under only traffic-length leakage is not a property of the public-receipt profile. Theorem~\ref{thm:confidentiality} states the guarantee with the actual exported metadata; payloads and provider secrets are not exposed through a plaintext forwarding interface.

\subsection{Restart, Replacement, and Update Fences}

A runtime incarnation is identified cryptographically by its measurement and channel/signing public keys. A service generation is management state associated with that incarnation; it does not replace evidence for the keys. Let $f=(a,b,g,v,t)$ identify a credential registration, binding, generation, configuration revision, and token revision. Within one incarnation, accepted updates advance a revision or install a fresh binding; retired fence tuples are not reused.

\begin{proposition}[Incarnation separation and local update consistency]
\label{prop:lifecycle}
Under $\mathsf{Good}$, redirecting an established session to a replacement with fresh private keys cannot continue that session. A restarted empty credential store cannot serve an account before successful provisioning. Within one live store, an update conditioned on a retired fence cannot overwrite a newer state when comparison and application are atomic.
\end{proposition}
\begin{proof}
The replacement does not hold the established channel keys. It must therefore obtain a fresh attested session; accepting old-channel records as that session would violate peer or record authentication, or require context reuse. Signing keys are checked against the original attestation, so a replacement close-out also cannot finish the old request. The restarted store's admission predicate requires an accepted credential binding, which is absent until provisioning installs it. Finally, an atomic update applies only if the stored fence equals its expected fence. Once the binding or relevant revision advances, that equality fails for an older operation. A duplicate completion can reuse its retained result without applying a second state transition.
\end{proof}

Planned replacement can drain old requests while admitting new ones on the new incarnation. A surviving old enclave retains its own keys until it stops; withdrawing a scheduler entry does not erase them. Revocation therefore governs subsequent authorization decisions, and provider-side token revocation terminates the provider credential itself. Historical records remain verifiable against their retained authorization snapshots.

\subsection{Composition and Implementation Obligations}

Define $\mathsf{BadAccept}$ as an honest client's confirmation of a result with any of: no corresponding runtime-issued record, a different retained request, an unauthorized execution trace, different completed response bytes, or a charge inconsistent with Eq.~\eqref{eq:confirmed-charge}.

\begin{corollary}[End-to-end confirmed-result integrity]
\label{cor:composition}
For the specified protocol machine,
\begin{equation}
 \Pr[\mathsf{BadAccept}]\leq\epsilon_{\rm auth}(\kappa).
\label{eq:composition-bound}
\end{equation}
Under the same model, the confidentiality guarantee is Eq.~\eqref{eq:confidentiality-bound} with leakage $\mathcal L$.
\end{corollary}
\begin{proof}
On $\mathsf{Good}$, Lemma~\ref{lem:record} gives record and request correspondence; Theorem~\ref{thm:execution} gives authorized invocation; Theorem~\ref{thm:stream} gives stream equality and successful closure; and Theorem~\ref{thm:charge} gives the bound usage-based charge. None of the alternatives in $\mathsf{BadAccept}$ can occur. Equation~\eqref{eq:auth-budget} bounds the probability of leaving $\mathsf{Good}$. The confidentiality statement is Theorem~\ref{thm:confidentiality}.
\end{proof}

The composition applies to executions of the specified machine. Its implementation obligations are agreement on typed parsing and serialization, signed request/channel context binding, fresh acceptance contexts, admission and trace coverage of every inference attempt including retries, and signing only after validated closure. In particular, fresh public nonces alone do not replace authenticated channel-instance binding, and retaining a received receipt digest does not replace computing the digest of the received stream. These are the refinement conditions connecting a concrete interpreter to the model. The functional appendix checks representative behavior; the theorems establish the protocol properties under the stated conditions.

Suppression can leave a valid authenticated prefix without a confirmed result. The protocol provides per-execution authenticity, not availability, log completeness, or financial fair exchange. Applications requiring a complete transaction gate irreversible actions on confirmation. These boundaries follow directly from the adversary's ability to drop messages and do not weaken the integrity of a result that is confirmed.

\begin{table*}[t]
\caption{Security conclusions and their scope in the protocol model.}
\label{tab:security-summary}
\centering
\small
\begin{tabular}{p{0.18\textwidth}p{0.45\textwidth}p{0.27\textwidth}}
\toprule
Property & Guarantee & Argument \\
\midrule
Authorized execution & Every issued attempt respects both authorization inputs, including initial route, enabling failure, endpoint, and attempt bound. & Proposition~\ref{prop:narrowing}, Lemma~\ref{lem:prefix}, Theorem~\ref{thm:execution} \\
Request and stream binding & Confirmation identifies the retained request and the complete runtime-emitted application bytes; replay cannot create a second live confirmation. & Lemma~\ref{lem:record}, Theorem~\ref{thm:stream} \\
Charge integrity & The confirmed amount uses the successful route, runtime-observed provider usage, and retained authorized pricing. & Theorem~\ref{thm:charge} \\
Channel confidentiality & Protected channels add no distinguishable leakage beyond exported interfaces, including public payload digests. & Theorem~\ref{thm:confidentiality} \\
Lifecycle separation & Fresh-key replacement cannot inherit an established session; stale updates cannot overwrite a newer live-store fence. & Proposition~\ref{prop:lifecycle} \\
\bottomrule
\end{tabular}
\end{table*}

\section{Implementation and Runtime Lifecycle}
\label{sec:implementation}

We implemented the client adapter, business-plane relay, attested runtime, shared contract engine, and verifier in Rust. The client adapter exposes an OpenAI-style streaming interface so an application can retain its existing API configuration. The relay performs coarse authentication and quota checks, distributes the registry, forwards opaque frames, and stores close-out records. The \amr{} owns provider credentials, request derivation, provider TLS, response framing, usage parsing, charge computation, and record signing.

The channel uses a Noise authenticated key exchange~\cite{noiseprotocol} followed by AEAD-protected request and response frames. Contracts and records use typed JSON schemas with deterministic serialization for commitments. Commitments use domain-separated SHA-256, signatures use Ed25519, and charge computation uses exact decimal values serialized as strings. The shared verifier implements policy intersection, trace validation, endpoint predicates, usage normalization, charge recomputation, and stream confirmation. The streaming adapter parses server-sent events incrementally and releases frames as they arrive.

The request-body commitment is computed after the runtime selects an initial route and derives its provider-facing model field. Client confirmation checks this commitment against the retained request and that initial route in the validated trace. Each fallback call is derived from the same retained request under its selected route. Upstream URL overrides are checked before network access against the route's scheme, host, port, and path prefix; an operator allowlist may restrict this set further. Response usage is extracted from streamed provider events and is mandatory for charge-bearing close-out.

The TEE interface is substrate-independent; the evaluated backend validates AWS Nitro Enclaves COSE attestation documents, certificate chains, PCR measurements, challenges, and key bindings~\cite{rfc9052,awsattestation,awsattestationvalidate}. The contract engine supports multi-attempt traces. The live-provider deployment used single-route contracts, while guarded fallback and failure outcomes were exercised through the networked mediation path described next.

\subsection{Credential Provisioning}

The credential owner and the inference client have different roles even when operated by the same party. The owner authorizes use of a provider account; the inference client authorizes a request contract. A registration identifier in the business plane is not a provider secret. Credentials enter the runtime in an owner-originated encrypted envelope after the owner has validated the registry, fresh attestation, and the runtime's channel key. The business plane stages and forwards this envelope but has no decryption key. Only the runtime opens it and uses the credential for provider authentication over TLS.

The complete implementation separates preparation, encrypted upload, verification, and activation. An upload contains a registration identifier, configuration revision, token revision, binding identifier, and runtime generation. These fields are included in the protected payload and checked against the selected operation. Successful forwarding alone is not completion: activation requires runtime acceptance for the same binding and revisions. A lost reply is reconciled against that operation rather than treated as evidence that the upstream credential was rejected.

Algorithm~\ref{alg:provision} summarizes the trust boundary. The owner retains its authoritative copy outside the gateway business plane. The runtime's credential repository is an in-memory store; restarting the process requires a new attested upload rather than recovering provider secrets from an operator-mounted file.

\begin{algorithm}[t]
\caption{Attested credential provisioning}
\label{alg:provision}
\footnotesize
\begin{algorithmic}[1]
\Require owner credential $K$, registration $a$, requested runtime
\State obtain fresh evidence binding runtime channel and signing keys
\State owner verifies registry, measurement, challenge, and key binding
\State establish the owner--runtime protected channel
\State prepare binding $b$, generation $g$, revisions $v,t$
\State seal $(K,a,b,g,v,t)$ to the verified runtime key
\State business plane forwards the sealed envelope
\State runtime opens envelope and checks the binding and revisions
\State runtime verifies the credential through the provider adapter
\State activate only the matching accepted binding
\State owner records acceptance for $(a,b,g,v,t)$
\end{algorithmic}
\end{algorithm}

\subsection{Rotation, Restart, and Replacement}

\textbf{Provider credential updates.} Configuration and token revisions distinguish a configuration change from replacement of secret material. The runtime rejects stale revisions and conflicting reuse of a binding. OAuth refresh reserves an operation against the loaded binding, expected token revision, and owner synchronization key; application of the result rechecks these values. The refreshed material is encrypted for the owner's synchronization key before crossing the business plane. Duplicate completion returns the retained operation result, and a delayed result cannot overwrite a newer loaded revision. Revision checks apply to the live in-memory state.

\textbf{Runtime identity.} Each runtime process generates separate channel and close-out signing keypairs in memory. Runtime registration carries an incarnation identifier; replacing the incarnation advances the service's generation and invalidates its old credential inventory for new admissions. Generation checks reject stale management operations, while attestation and channel authentication establish the cryptographic identity.

\textbf{Restart recovery.} After a process restart, the new process has fresh private keys and an empty credential store. Clients and credential owners establish new attested sessions. Accounts become available on that instance after matching credentials have been uploaded and accepted. An interrupted execution retains its original acceptance context and is unconfirmed if no valid close-out was received.

\textbf{Planned replacement.} The replacement process starts under an authorized measurement, proves its fresh keys, and receives fresh encrypted uploads. Once ready, new admissions can move to it while existing requests on the old instance drain under their original session and signing keys. An established session is not transplanted merely because the new instance has the same measurement. Stored records retain the verification context of their issuing instance. Updating a credential or withdrawing an instance from scheduling does not remotely erase an old live enclave or revoke a provider token; provider-side revocation is needed when that secret itself must cease to work.

\section{Evaluation}
\label{sec:evaluation}

We evaluate how the protected path scales with concurrency and response shape, decompose its software cost, and examine the end-to-end Nitro deployment. The three-path comparison measures the complete mechanism: channel setup, encryption, contract handling, stream commitments, signing, and client confirmation. Appendix~\ref{app:functional} documents functional scenario coverage.

\subsection{Evidence Sources and Experimental Topology}

The software-path experiments ran as release builds on one Apple M4 Pro host with 14 logical CPUs, Darwin 24.5.0 arm64, and Rust 1.95.0. The client driver, gateway path, and controlled OpenAI-style endpoint run on that host. This topology holds response content and event timing fixed to expose mediation cost. The controlled endpoint emits server-sent events (SSE): incremental HTTP response events containing application chunks, usage, and completion markers. These measurements include the implemented channel, framing, commitment, signing, and verification work; they do not include Nitro isolation or host--enclave transport overhead.

The external deployment places the runtime in AWS Nitro Enclaves, with an untrusted parent-host relay and forwarding bridge. The client adapter reaches the relay over HTTP/gRPC; opaque client--runtime frames cross the host--enclave bridge, and the runtime establishes provider TLS through that forwarding path. The retained deployment record identifies an enclave allocation of two vCPUs and 512~MiB. These records establish the attestation and live execution path; matched latency and throughput comparisons use the fully specified single-host topology above.

For latency experiments, each trial produces a p95 over its measured requests. Figures report the median of trial-level p95 values; throughput is computed from trial wall-clock time. Protected-minus-plaintext differences are paired within the same trial. Shaded intervals, where shown, are interquartile ranges across paired trial values. The trial is the unit used to summarize measurement variability.

\subsection{What the Contract Adds}

Table~\ref{tab:example} identifies the semantic distinction: destination membership admits both $A$ and $B$, while the execution predicate restricts when $B$ may be used. Proposition~\ref{prop:narrowing} establishes that policy intersection preserves this restriction. The networked workload exercises the implementation through provider-shaped API responses and controlled failures.

Permitted fallback produced an accepted record containing the primary failure and subsequent success. In the hidden-fallback case, removing the primary attempt while retaining the fallback result caused semantic rejection. The response remained available, but its evidence lacked the history required by the contract. This case illustrates the execution predicate; Appendix~\ref{app:functional} lists the complete scenario coverage.

\subsection{Latency, Throughput, and Response Shape}

We compare direct access to the instrumented SSE endpoint, a plaintext gateway, and the contract-aware gateway. All paths receive the same request and return eight 512-B response chunks, a usage object, and a completion event on the same schedule. Concurrency ranges from 1 to 32. Each path receives 100 warm-up and 1,000 measured requests in each of 30 trials, with path order permuted per trial. The plaintext path provides a forwarding baseline; the protected-minus-plaintext difference measures the added cost of the complete protected mechanism.

\begin{figure*}[t]
  \centering
  \includegraphics[width=0.96\textwidth]{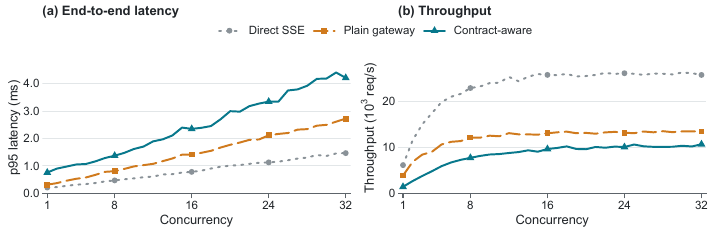}
  \caption{Concurrency scaling. Lines are medians across 30 trials at every integer concurrency; markers identify 1/8/16/24/32. Each point summarizes 1,000 requests per path and trial.}
  \Description{Two plots compare p95 latency and throughput for direct SSE, plaintext gateway, and contract-aware paths at concurrency levels from one to thirty-two.}
  \label{fig:concurrency}
\end{figure*}

Figure~\ref{fig:concurrency} shows trial-median p95 latency and throughput. At concurrency one, the protected path added 0.442~ms paired p95 over the plaintext gateway. At concurrency 32, p95 latency was 4.215~ms versus 2.727~ms. The separately observed throughput peaks were 10.65 and 13.50 thousand requests/s for the protected and plaintext paths, respectively, a ratio of 78.9\%. At concurrency 16, the first decrypted frame arrived at 1.520~ms p95, before close-out. The path therefore retained incremental delivery while performing final verification at stream completion.

At concurrency 32, the interval from the last decrypted frame through record verification was 379~\micros{} p95. Signing and client verification took 107 and 143~\micros{} p95, respectively, and record size was 1,748~B p95. The analysis includes 2,880,000 measured requests from the concurrency experiment and 96,000 from the selected response-shape sweeps, excluding warm-up. These are controlled endpoint invocations, not live-provider requests. The retained measurements contain no request error, and protected responses completed client confirmation.

\begin{figure*}[t]
  \centering
  \includegraphics[width=0.96\textwidth]{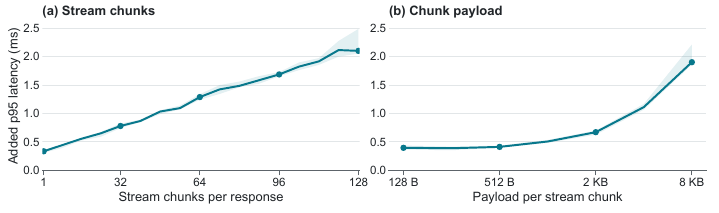}
  \caption{Response-shape sensitivity. Curves show the median paired difference between contract-aware and plaintext-gateway p95 latency across ten trials; shading denotes the interquartile range.}
  \Description{Two plots show protected-path latency overhead increasing with stream chunk count and payload bytes per chunk. Shaded bands show variability across trials.}
  \label{fig:shape}
\end{figure*}

Figure~\ref{fig:shape} characterizes how protected-path overhead changes with response shape. We vary chunk count from 1 to 128 at 512~B each and chunk payload from 128~B to 8~KB at eight chunks. Each cell contains 200 requests in each of ten trials. Increasing chunk count from 1 to 128 raised the median paired p95 overhead from 0.336 to 2.101~ms. Increasing chunk payload from 128~B to 8~KB raised it from 0.397 to 1.902~ms. Thus, overhead increased with both the number and size of streamed chunks.

\subsection{Cost Decomposition}

The response-shape curves are consistent with the work performed by the streaming path: each additional chunk incurs framing, authenticated encryption, and commitment updates, while larger chunks increase byte-processing cost. Retained cryptographic microbenchmarks quantify these components independently of endpoint scheduling. For 512-B chunks, channel setup took 145.625~\micros{} p95, response-chunk encryption/decryption took 2.344~\micros{} p95 per chunk, and commitment updates took 1.120~\micros{} p95 per chunk. The full-path signing, verification, and close-out measurements above capture the terminal work.

These component measurements explain why both chunk count and payload size affect overhead. They also distinguish per-session setup, per-chunk work, and final verification. Their percentiles are measured separately and are not summed to reconstruct an end-to-end percentile.

\subsection{Attested Cloud Deployment}

The Nitro verifier checked the attestation document and certificate chain, fresh challenge, PCR measurement, and binding of channel and signing keys. Attestation verification and encrypted-session setup took 4.849 and 4.001~ms p95, respectively, as once-per-session costs. Seven attestation and record-binding scenarios are documented in Appendix~\ref{app:functional}.

\begin{table}[t]
\caption{Nitro deployment measurements. Setup costs are p95; live requests used one authorized route per contract.}
\label{tab:nitro}
\centering
\footnotesize
\setlength{\tabcolsep}{3.3pt}
\begin{tabular}{lr}
\toprule
Measurement & Result \\
\midrule
Attestation verification & 4.849 ms p95 \\
Encrypted-session setup & 4.001 ms p95 \\
Live streaming requests & 30 completed \\
Request concurrency & 1, 4, 8 \\
\bottomrule
\end{tabular}
\end{table}

The deployed path carried 30 live streaming LLM API requests using the recorded model identifier \texttt{gpt-5.5}, at concurrency 1, 4, and 8. Each returned a completed stream, provider-reported usage, a response identifier, and a signed close-out accepted by the client. Relay storage contained no request body, response body, or provider credential. These executions connect the client adapter, opaque relay, attested runtime, provider TLS, and final confirmation in one deployed single-route path. Multi-route behavior is covered by the controlled networked workload.

\subsection{User-Visible Latency}

Millisecond-scale mediation adds little latency to upstream-bound inference. Across the live-provider concurrency settings, median time to first response byte ranged from 1.293 to 1.330~s, and median request completion time ranged from 3.296 to 3.383~s. These seconds-scale waits place gateway processing on a substantially shorter timescale than the remote inference path. Content is forwarded as it arrives, and close-out verification follows the final content frame rather than delaying stream delivery. Thus, the measured software overhead is a small latency cost for this workload, while the controlled throughput curves characterize the gateway's processing capacity.

\section{Discussion}

\textbf{Late binding and deployment.} Changes to the interpreter require a new authorized measurement. Changes to available routes, fallback guards, endpoints, and rates are authenticated policy inputs. This separation gives clients stable enforcement semantics while allowing operators to maintain a changing provider catalog. Reproducible builds or source-to-measurement audits establish which implementation a client authorizes; the signed registry distributes that authorization without replacing the client's local measurement policy.

\textbf{Replication and multi-tenancy.} The business plane schedules new sessions across replicas. Each replica proves its measurement and keys, and each established channel identifies its cryptographic peer. Clients can supply different provider sets and fallback guards to the same measured interpreter, supporting tenant-specific authorization without per-tenant images. Instance replacement follows the provisioning and draining protocol in Section~\ref{sec:implementation}.

\textbf{Execution evidence and settlement.} The runtime enforces authorization before and during provider invocation. The close-out makes the chosen branch, observed usage, and completed stream portable for client confirmation and later rechecking. Settlement can compare an invoice with the signed expected charge under the original pricing policy. The protocol authenticates those inputs; receipt delivery and financial fair exchange remain separate from per-execution authenticity.

\section{Related Work}

\textbf{Cloud gateways and service routing.} Commercial and open-source LLM gateways provide multi-provider routing, fallback, API normalization, and spend controls~\cite{openrouter_routing,openrouter_fallback,litellm_gateway,cloudflare_dynamic_routing,portkey_fallbacks}. Work on service-oriented computing similarly models service selection and quality obligations~\cite{papazoglou2003soc,keller2003wsla,liu2004qos}. These systems motivate our contract fields but generally trust the gateway operator to execute and report routing decisions.

\textbf{Confidential cloud execution.} TEEs and systems including Haven and SCONE isolate cloud workloads from privileged hosts~\cite{baumann2014haven,arnautov2016scone,sok2022tee}. Remote attestation authenticates measured software and key material~\cite{coker2011principles,rfc9334}. We use these primitives to place request-level contract interpretation and provider invocation within one attested boundary.

\textbf{Protected LLM intermediaries.} Portcullis protects selected fields inside an attested third-party LLM gateway~\cite{portcullis2025}. OpenGradient's TEE Gateway performs multi-provider request transformation, response signing, and usage-based payment inside Nitro Enclaves~\cite{opengradient2026}. AEX signs request--output relations and stream commitments across trusted issuers~\cite{guan2026aex}. Our focus is the client's authorization of a complete mediation path: initial routes, failure-guarded transitions, endpoints, pricing, and delivered stream. The effective contract governs execution, and the close-out retains its trace and charge derivation for verification.

\section{Conclusion}

Evidence-bound gateway-path provenance connects client authorization to an attested execution and its delivered result. Policy intersection and the runtime state machine preserve initial-route, endpoint, and guarded-fallback constraints. Signed context and stream binding connect confirmation to one request, its observed usage, and its authorized charge. The security analysis composes these properties and makes the leakage of public execution metadata explicit. Credential provisioning and fresh-key replacement preserve the runtime boundary across lifecycle changes. Controlled measurements characterize the software cost across concurrency and response shapes, while the Nitro prototype demonstrates live single-route execution. The resulting architecture separates gateway administration from trusted payload processing and supplies a verifiable account of confirmed mediation.

\appendix
\section{Functional Scenario Coverage}
\label{app:functional}

The validation artifact retains inputs, expected decisions, execution records, and verifier outcomes. Table~\ref{tab:coverage} counts distinct scenarios within each workload. The lists below specify their obligations and observed outcomes; repeated executions of a fixed case are not additional scenarios. These cases check implementation behavior against the protocol defined in Sections~\ref{sec:model}--\ref{sec:security}.

\begin{table}[t]
\caption{Functional coverage by distinct scenario. Attestation checks form a separate workload from the four mediation groups.}
\label{tab:coverage}
\centering
\footnotesize
\begin{tabular}{lr}
\toprule
Workload & Scenarios \\
\midrule
Request and stream binding & 14 \\
Usage binding & 7 \\
Charge derivation & 6 \\
Networked route and fallback & 7 \\
Attestation and key binding & 7 \\
\bottomrule
\end{tabular}
\end{table}

\subsection{Request and Stream Binding}

The valid record was accepted. The following modifications were rejected: a missing record; a modified record; a signing key different from the attested key; a record swapped across requests; a wrong provider, model, or route label; a mismatched stream commitment; deletion, reordering, substitution, or duplication of a chunk; and truncation despite delivery of a terminal record. These scenarios exercise authorship, request context, and equality with the client-observed stream.

\subsection{Usage and Charge Binding}

The usage workload accepted valid usage evidence and rejected six variants: absent provider usage, disagreement between record usage and the observed provider usage, modification after signing, cross-request substitution, a wrong provider, and a wrong endpoint. The charge workload accepted the valid charge and rejected a modified amount, a wrong pricing-policy version or commitment, a wrong provider/model rate, substituted usage, and an invoice inconsistent with the signed charge. Invoice comparison occurs at settlement; the other charge checks belong to final acceptance.

\subsection{Networked Route and Fallback}

The API/SSE workload connects controlled provider responses to the policy engine and final verifier. Primary success and permitted fallback were accepted. A disallowed provider, an endpoint-policy violation, and a failed primary with fallback disabled were rejected. Removing the primary attempt from a fallback trace caused semantic rejection by the final verifier; modifying a signed record caused signature rejection. The latter cases deliberately allow a provider-shaped response to complete before auditing its evidence, thereby testing the final-acceptance boundary independently of response availability.

\subsection{Attestation and Key Binding}

The Nitro workload accepted a valid execution and rejected record tampering, a wrong signing key, stream substitution, a stale challenge, a wrong runtime measurement, and a registry mismatch. These checks cover the link from the client challenge through the accepted measurement and keys to the returned record and stream.

\subsection{Business-Plane Visibility}

The isolation check inserts distinct canaries into the prompt, response, and provider credential and scans the captured bytes available to the business plane. The plaintext gateway exposed all three markers. The protected path exposed none of them in its captured request/response envelopes. The observation checks payload handling at the implemented interfaces; the cryptographic isolation argument is given in Section~\ref{sec:security}.

\bibliographystyle{ACM-Reference-Format}
\bibliography{references}

\end{document}